\documentclass[english]{article}
\usepackage[T1]{fontenc}
\usepackage[latin9]{inputenc}
\usepackage{amsmath}
\usepackage{amssymb}
\usepackage{graphicx}

\makeatletter
\renewcommand{\vec}[1]{\mathbf{#1}}
\usepackage{url}
\usepackage{cite}
\renewcommand{\Re}{\operatorname{Re}}
\renewcommand{\Im}{\operatorname{Im}}

\date{Created August 2007; updated \today}

\makeatother

\usepackage{babel}
\begin{document}
\title{Notes on Perfectly Matched Layers (PMLs)}
\author{Steven G. Johnson}
\maketitle
\begin{abstract}
This note is intended as a brief introduction to the theory and practice
of perfectly matched layer (PML) absorbing boundaries for wave equations, originally developed for MIT courses 18.369 and 18.336. It
focuses on the complex stretched-coordinate viewpoint, and also discusses
the limitations of PML.
\end{abstract}
\tableofcontents{}

\section{Introduction}

Whenever one solves a PDE numerically by a volume discretization,\footnote{As opposed to a boundary discretization, e.g. in boundary-element
methods, where the unknowns are on the interfaces between homogeneous
regions, and the homogeneous regions are handled analtyically. In
this case, no artificial truncation is required...except in the case
of interfaces that extend to infinity, which lead to some interesting
unsolved problems in boundary-element methods.} one must truncate the computational grid in some way, and the key
question is how to perform this truncation without introducing significant
artifacts into the computation. Some problems are naturally truncated,
e.g. for periodic structures where periodic boundary conditions can
be applied. Some problems involve solutions that are rapidly decaying
in space, so that the truncation is irrelevant as long as the computational
grid is large enough. Other problems, such as Poisson's equation,
involve solutions that vary more and more slowly at greater distances---in
this case, one can simply employ a coordinate transformation, such
as $\tilde{x}=\tanh x$, to remap from $x\in(-\infty,\infty)$ to
$\tilde{x}\in(-1,1)$, and solve the new finite system. However, some
of the most difficult problems to truncate involve \emph{wave equations},
where the solutions are \emph{oscillating} and typically decay with
distance $r$ only as $1/r^{(d-1)/2}$ in $d$ dimensions.\footnote{The square of the solutions are typically related to a rate of energy
transport, e.g. the Poynting vector in electromagnetism, and energy
conservation requires that this decay be proportional to the surface
area $\sim r^{d-1}$.} The slow decay means that simply truncating the grid with hard-wall
(Dirichlet or Neumann) or periodic boundary conditions will lead to
unacceptable artifacts from boundary reflections. The oscillation
means that any real coordinate remapping from an infinite to a finite
domain will result in solutions that oscillate infinitely fast as
the boundary is approached---such fast oscillations cannot be represented
by any finite-resolution grid, and will instead effectively form a
reflecting hard wall. Therefore, wave equations require something
different: an \emph{absorbing boundary} that will somehow absorb waves
that strike it, without reflecting them, and without requiring infeasible
resolution.

The first attempts at such absorbing boundaries for wave equations
involved \emph{absorbing boundary conditions} (\emph{ABC}s) \cite{Taflove00}.
Given a solution on a discrete grid, a \emph{boundary condition} is
a rule to set the value at the edge of the grid. For example, a simple
Dirichlet boundary condition sets the solution to zero at the edge
of the grid (which will reflect waves that hit the edge). An ABC tries
to somehow \emph{extrapolate} from the interior grid points to the
edge grid point(s), to fool the solution into ``thinking'' that
it extends forever with no boundary. It turns out that this is possible
to do perfectly in one dimension, where waves can only propagate in
two directions ($\pm x$). However, the main interest for numerical
simulation lies in two and three dimensions, and in these cases the
infinite number of possible propagation directions makes the ABC problem
much harder. It seems unlikely that there exists any efficient method
to exactly absorb radiating waves that strike a boundary at any possible
angle. Existing ABCs restrict themselves to absorbing waves exactly
only at a few angles, especially at normal incidence: as the size
of the computational grid grows, eventually normal-incident waves
must become the dominant portion of the radiation striking the boundaries.
Another difficulty is that, in many practical circumstances, the wave
medium is not homogeneous at the grid boundaries. For example, to
calculate the transmission around a dielectric waveguide bend, the
waveguide (an inhomogeneous region with a higher index of refraction)
should in principle extend to infinity before and after the bend.
Many standard ABCs are formulated only for homogeneous materials at
the boundaries, and may even become numerically unstable if the grid
boundaries are inhomogeneous.

\begin{figure}
\begin{centering}
\includegraphics[width=1\columnwidth]{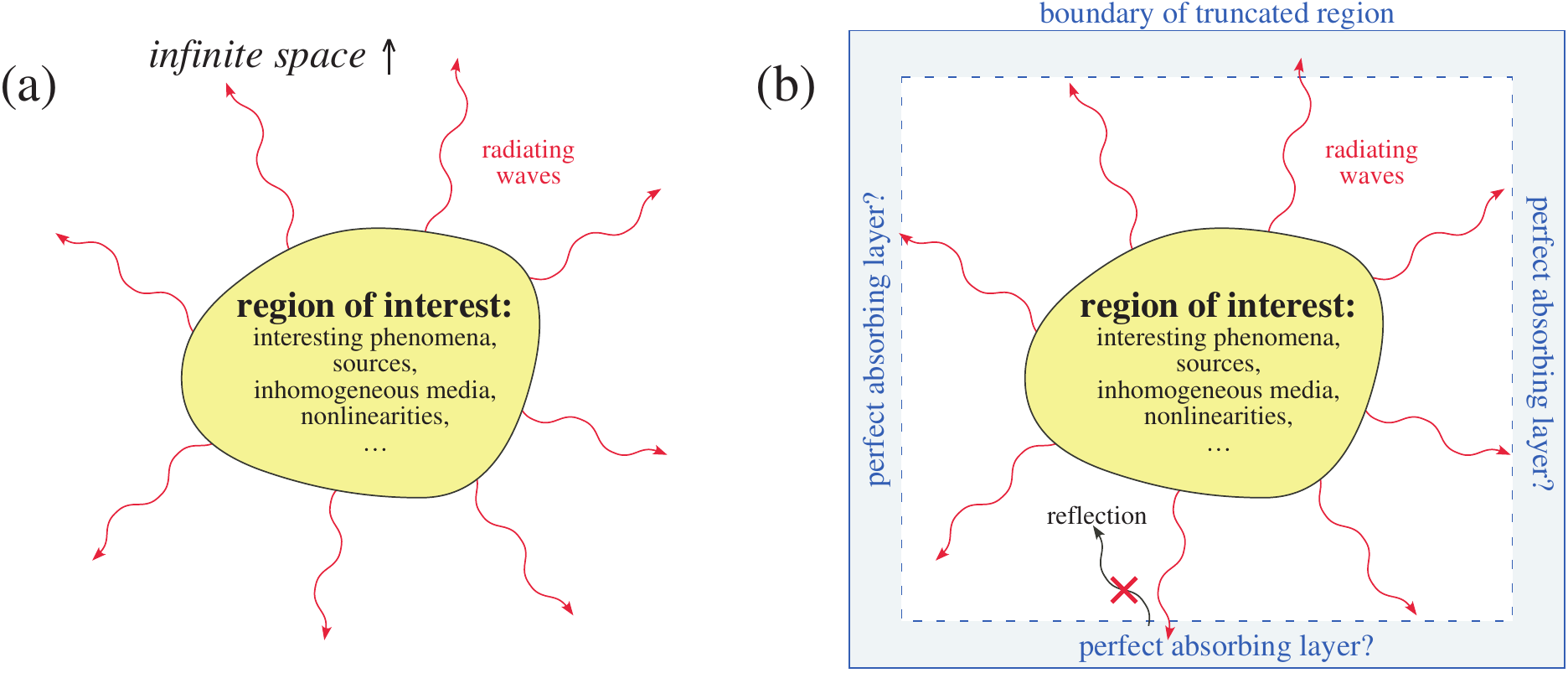}
\par\end{centering}
\caption{\label{fig:absorbing-layer}(a) Schematic of a typical wave-equation
problem, in which there is some finite region of interest where sources,
inhomogeneous media, nonlinearities, etcetera are being investigated,
from which some radiative waves escape to infinity. (b) The same problem,
where space has been truncated to some computational region. An absorbing
layer is placed adjacent to the edges of the computational region---a
\emph{perfect} absorbing layer would absorb outgoing waves without
reflections from the edge of the absorber.}
\end{figure}
In 1994, however, the problem of absorbing boundaries for wave equations
was transformed in a seminal paper by Berenger \cite{Berenger94}.
Berenger changed the question: instead of finding an absorbing boundary
\emph{condition}, he found an absorbing boundary \emph{layer}, as
depicted in Fig\@.~\ref{fig:absorbing-layer}. An absorbing boundary
\emph{layer} is a layer of artificial absorbing material that is placed
adjacent to the edges of the grid, completely \emph{independent of
the boundary condition}. When a wave enters the absorbing layer, it
is attenuated by the absorption and decays exponentially; even if
it reflects off the boundary, the returning wave after one round trip
through the absorbing layer is exponentially tiny. The problem with
this approach is that, whenever you have a transition from one material
to another,\footnote{Technically, reflections occur when translational symmetry is broken.
In a periodic structure (discrete translational symmetry), there are
waves that propagate without scattering, and a uniform medium is just
a special case with period $\rightarrow0$.} waves generally reflect, and the transition from non-absorbing to
absorbing material is no exception---so, instead of having reflections
from the grid boundary, you now have reflections from the absorber
boundary. However, Berenger showed that a special absorbing medium
could be constructed so that waves do \emph{not} reflect at the interface:
a \emph{perfectly matched layer}, or \emph{PML}. Although PML was
originally derived for electromagnetism (Maxwell's equations), the
same ideas are immediately applicable to other wave equations.

There are several equivalent formulations of PML. Berenger's original
formulation is called the \emph{split-field} PML, because he artificially
split the wave solutions into the sum of two new artificial field
components. Nowadays, a more common formulation is the \emph{uniaxial}
PML or \emph{UPML}, which expresses the PML region as the ordinary
wave equation with a combination of artificial \emph{anisotropic}
absorbing materials \cite{Sacks95}. Both of these formulations were
originally derived by laboriously computing the solution for a wave
incident on the absorber interface at an arbitrary angle (and polarization,
for vector waves), and then solving for the conditions in which the
reflection is always zero. This technique, however, is labor-intensive
to extend to other wave equations and other coordinate systems (e.g.
cylindrical or spherical rather than Cartesian). It also misses an
important fact: PML still works (can still be made theoretically reflectionless)
for \emph{inhomogeneous} media, such as waveguides, as long as the
medium is homogeneous in the direction perpendicular to the boundary,
even though the wave solutions for such media cannot generally be
found analytically. It turns out, however, that \emph{both} the split-field
and UPML formulations can be derived in a much more elegant and general
way, by viewing them as the result of a \emph{complex coordinate stretching}
\cite{Chew94,Rappaport95,Teixeira98}.\footnote{It is sometimes implied that only the split-field PML can be derived
via the stretched-coordinate approach \cite{Taflove00}, but the UPML
media can be derived in this way as well \cite{Teixeira98}.} It is this complex-coordinate approach, which is essentially based
on \emph{analytic continuation} of Maxwell's equations into complex
spatial coordinates where the fields are exponentially decaying, that
we review in this note.

In the following, we first briefly remind the reader what a wave equation
is, focusing on the simple case of the scalar wave equation but also
giving a general definition. We then derive PML as a combination of
two steps: analytic continuation into complex coordinates, then a
coordinate transformation back to real coordinates. Finally, we discuss
some limitations of PML, most notably the fact that it is no longer
reflectionless once the wave equation is discretized, and common workarounds
for these limitations.

\section{Wave equations}

There are many formulations of waves and wave equations in the physical
sciences. The prototypical example is the (source-free) \emph{scalar
wave equation:} 
\begin{equation}
\nabla\cdot(a\nabla u)=\frac{1}{b}\frac{\partial^{2}u}{\partial t^{2}}=\frac{\ddot{u}}{b}\label{eq:scalar-wave}
\end{equation}
where $u(\vec{x},t)$ is the scalar wave amplitude and $c=\sqrt{ab}$
is the phase velocity of the wave for some parameters $a(\vec{x})$
and $b(\vec{x})$ of the (possibly inhomogeneous) medium. For lossless,
propagating waves, $a$ and $b$ should be real and positive.

Both for computational convenience (in order to use a staggered-grid
leap-frog discretization) and for analytical purposes, it is more
convenient to split this second-order equation into two coupled first-order
equation, by introducing an auxiliary field $\vec{v}(\vec{x},t)$:
\begin{eqnarray}
\frac{\partial u}{\partial t} & = & b\nabla\cdot\vec{v},\label{eq:scalar-wave-udot}
\end{eqnarray}
\begin{equation}
\frac{\partial\vec{v}}{\partial t}=a\nabla u,\label{eq:scalar-wave-vdot}
\end{equation}
which are easily seen to be equivalent to eq\@.~(\ref{eq:scalar-wave}).

Equations~(\ref{eq:scalar-wave-udot}--\ref{eq:scalar-wave-vdot})
can be written more abstractly as:

\begin{equation}
\frac{\partial\vec{w}}{\partial t}=\frac{\partial}{\partial t}\left(\begin{array}{c}
u\\
\vec{v}
\end{array}\right)=\left(\begin{array}{cc}
 & b\nabla\cdot\\
a\nabla
\end{array}\right)\left(\begin{array}{c}
u\\
\vec{v}
\end{array}\right)=\hat{D}\vec{w}\label{eq:scalar-wave-abstract}
\end{equation}
for a $4\times4$ linear operator $\hat{D}$ and a 4-component vector
$\vec{w}=(u;\vec{v})$ (in three dimensions). The key property that
makes this a ``wave equation'' turns out to be that $\hat{D}$ is
an \emph{anti-Hermitian} operator in a proper choice of inner product,
which leads to oscillating solutions, conservation of energy, and
other ``wave-like'' phenomena. Every common wave equation, from
scalar waves to Maxwell's equations (electromagnetism) to Schrödinger's
equation (quantum mechanics) to the Lamé-Navier equations for elastic
waves in solids, can be written in the abstract form $\partial\vec{w}/\partial t=\hat{D}\vec{w}$
for some wave function $\vec{w}(\vec{x},t)$ and some anti-Hermitian
operator $\hat{D}$.\footnote{See e.g. Ref\@.~\cite{Johnson07-wave-equations}}
The same PML ideas apply equally well in all of these cases, although
PML is most commonly applied to Maxwell's equations for computational
electromagnetism.

\section{Complex coordinate stretching}

Let us start with the solution $\vec{w}(\vec{x},t)$ of some wave
equation in infinite space, in a situation similar to that in Fig\@.~\ref{fig:absorbing-layer}(a):
we have some region of interest near the origin $\vec{x}=0$, and
we want to truncate space outside the region of interest in such a
way as to absorb radiating waves. In particular, we will focus on
truncating the problem in the $+x$ direction (the other directions
will follow by the same technique). This truncation occurs in three
conceptual steps, summarized as follows:
\begin{enumerate}
\item In infinite space, \emph{analytically continue} the solutions and
equations to a \emph{complex} $x$ contour, which \emph{changes} oscillating
waves into \emph{exponentially decaying waves} outside the region
of interest \emph{without} reflections.
\item Still in infinite space, perform a \emph{coordinate transformation}
to express the complex $x$ as a function of a real coordinate. In
the new coordinates, we have \emph{real coordinates} and \emph{complex
materials}.
\item Truncate the domain of this new real coordinate inside the complex-material
region: since the solution is decaying there, as long as we truncate
it after a long enough distance (where the exponential tails are small),
it won't matter what boundary condition we use (hard-wall truncations
are fine).
\end{enumerate}
For now, we will make two simplifications:
\begin{itemize}
\item We will assume that the space far from the region of interest is \emph{homogeneous}
(deferring the inhomogeneous case until later). 
\item We will assume that the space far from the region of interest is linear
and time-invariant.
\end{itemize}
Under these assumptions, the radiating solutions in infinite space
must take the form of a superposition of \emph{planewaves}: 
\begin{equation}
\vec{w}(\vec{x},t)=\sum_{\vec{k},\omega}\vec{W}_{\vec{k},\omega}e^{i(\vec{k}\cdot\vec{x}-\omega t)},\label{eq:planewave-sum}
\end{equation}
for some constant amplitudes $\vec{W}_{\vec{k},\omega}$, where $\omega$
is the (angular) frequency and $\vec{k}$ is the wavevector. (In an
isotropic medium, $\omega$ and $\vec{k}$ are related by $\omega=c|\vec{k}|$
where $c(\omega)$ is some phase velocity, but we don't need to assume
that here.) In particular, the key fact is that the radiating solutions
may be decomposed into functions of the form 
\begin{equation}
\vec{W}(y,z)e^{i(kx-\omega t)}.\label{eq:planewave-x}
\end{equation}
The ratio $\omega/k$ is the phase velocity, which can be different
from the group velocity $d\omega/dk$ (the velocity of energy transport,
in lossless media). For waves propagating in the $+x$ direction,
the group velocity is positive. Except in very unusual cases, the
phase velocity has the same sign as the group velocity in a homogeneous
medium,\footnote{The formulation of PML absorbers when the phase velocity has sign
opposite to the group velocity, for example in the ``left-handed
media'' of electromagnetism, is somewhat more tricky~\cite{Cummer04,Dong04};
matters are even worse in certain waveguides with both signs of group
velocity at the same~$\omega$~\cite{LohOs09}.} so \textbf{we will assume} that $k$ is \textbf{positive}.

\subsection{Analytic continuation}

\begin{figure}
\begin{centering}
\includegraphics[width=1\columnwidth]{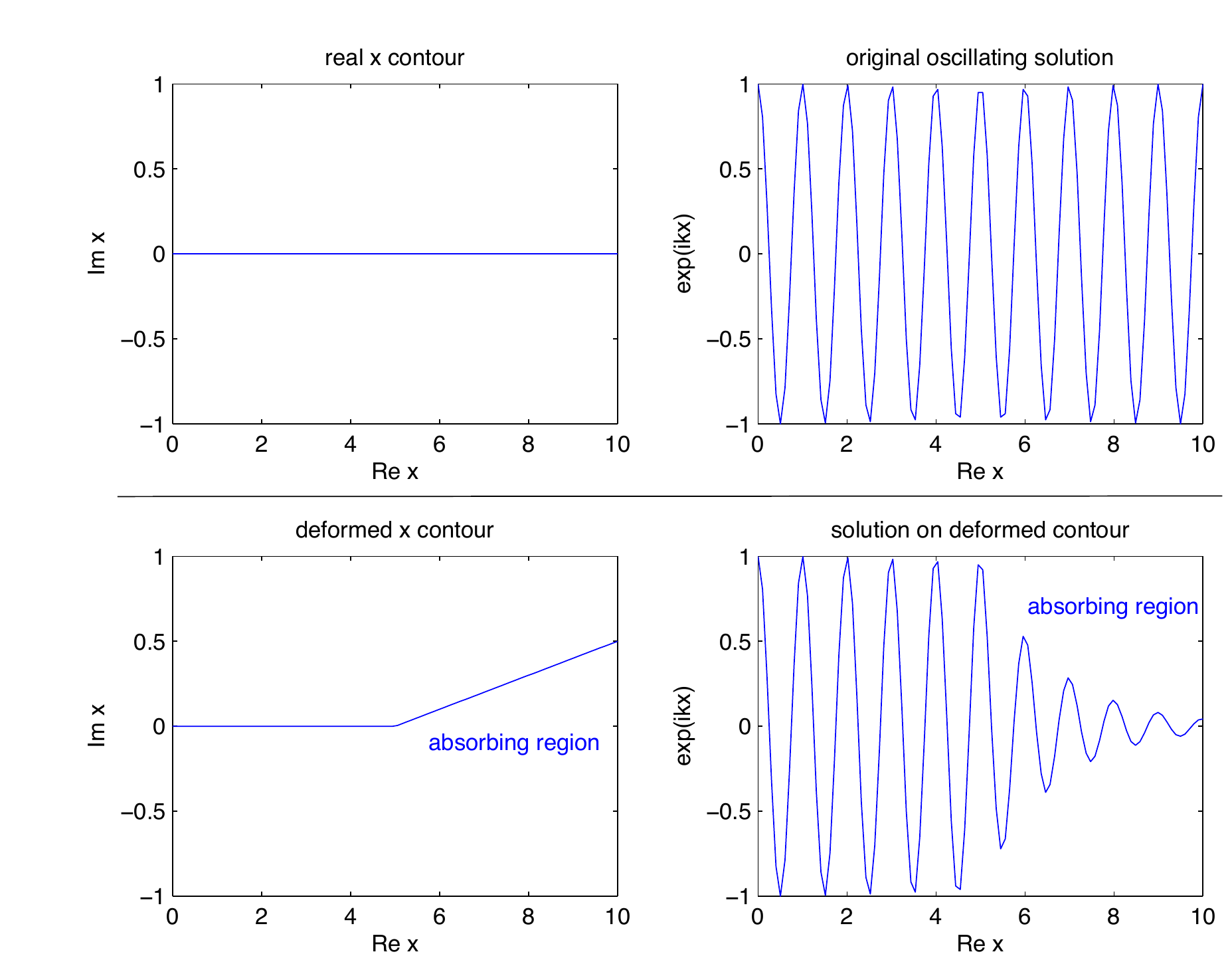}
\par\end{centering}
\caption{\emph{\label{fig:pml-continuation}Top}: real part of original oscillating
solution $e^{ikx}$ (right) corresponds to $x$ along the real axis
in the complex-$x$ plane (left). \emph{Bottom}: We can instead evaluate
this analytic function along a \emph{deformed} contour in the complex
plane: here (left) we deform it to increase along the imaginary axis
for $x>5$. The $e^{ikx}$ solution (right) is unchanged for $x<5$,
but is exponentially decaying for $x>5$ where the contour is deformed,
corresponding to an ``absorbing'' region.}
\end{figure}

The key fact about eq\@.~(\ref{eq:planewave-x}) is that it is an
\emph{analytic function} of $x$. That means that we can freely \emph{analytically
continue} it, evaluating the solution at \emph{complex values} of
$x$. The original wave problem corresponds to $x$ along the real
axis, as shown in the top panels of Fig\@.~\ref{fig:pml-continuation},
which gives an oscillating $e^{ikx}$ solution. However, if instead
of evaluating $x$ along real axis, consider what happens if we evaluate
it along the contour shown in the bottom-left panel of Fig\@.~\ref{fig:pml-continuation},
where for $\Re x>5$ we have added a linearly growing imaginary part.
In this case, the solution is \emph{exponentially} decaying for $\Re x>5$,
because $e^{ik(\Re x+i\Im x)}=e^{ik\Re x}e^{-k\Im x}$ is exponentially
decaying (for $k>0$) as $\Im x$ increases. That is, the solution
in this region acts like the solution in an \emph{absorbing material}.

However, there is one crucial difference here from an ordinary absorbing
material: the solution is \emph{not changed} for $\Re x<5$, where
$x$ is no different from before. So, it not only acts like an absorbing
material, it acts like a \emph{reflectionless absorbing material},
a PML.

The thing to remember about this is that the analytically continued
solution satisfies the \emph{same} differential equation. We assumed
the differential equation was $x$-invariant in this region, so $x$
only appeared in derivatives like $\frac{\partial}{\partial x}$,
and the derivative of an analytic function is the same along any $dx$
direction in the complex plane. So, we have succeeded in transforming
our original wave equation to one in which the radiating solutions
(large $|x$|) are exponentially decaying, while the part we care
about (small $x$) is unchanged. The only problem is that solving
differential equations along contours in the complex plane is rather
unfamiliar and inconvenient. This difficulty is easily fixed.

\subsection{Coordinate transformation back to real $x$}

For convenience, let's denote the complex $x$ contour by $\tilde{x}$,
and reserve the letter $x$ for the real part. Thus, we have $\tilde{x}(x)=x+if(x)$,
where $f(x)$ is some function indicating how we've deformed our contour
along the imaginary axis. Since the complex coordinate $\tilde{x}$
is inconvenient, we will just \emph{change variables} to write the
equations in terms of $x$, the real part!

Changing variables is easy. Whereever our original equation has $\partial\tilde{x}$
(the differential along the deformed contour $\tilde{x}$), we now
have $\partial\tilde{x}=(1+i\frac{df}{dx})\partial x$. That's it!
Since our original wave equation was assumed $x$-invariant (at least
in the large-$x$ regions where $f\neq0$), we have no other substitutions
to make. As we shall soon see, it will be convenient to denote $\frac{df}{dx}=\frac{\sigma_{x}(x)}{\omega}$,
for some function $\sigma_{x}(x)$. {[}For example, in the bottom
panel of Fig\@.~\ref{fig:pml-continuation}, we chose $\sigma_{x}(x)$
to be a step function: zero for $x\leq5$ and a positive constant
for $x>5$, which gave us exponential decay.{]} In terms of $\sigma_{x}$,
the entire process of PML can be conceptually summed up by a single
transformation of our original differential equation:
\begin{equation}
\boxed{\frac{\partial}{\partial x}\rightarrow\frac{1}{1+i\frac{\sigma_{x}(x)}{\omega}}\frac{\partial}{\partial x}.}\label{eq:pml-transformation}
\end{equation}
In the \textbf{PML regions} where $\sigma_{x}>0$, our oscillating
solutions turn into exponentially decaying ones. In the regions where
$\sigma_{x}=0$, our wave equation is unchanged and the \emph{solution
is unchanged}: there are no reflections because this is only an analytic
continuation of the original solution from $x$ to $\tilde{x}$, and
where $\tilde{x}=x$ the solution cannot change.

Why did we choose $\sigma_{x}/\omega$, as opposed to just $\sigma_{x}$?
The answer comes if we look at what happens to our wave $e^{ikx}$.
In the new coordinates, we get: 
\begin{equation}
e^{ikx}e^{-\frac{k}{\omega}\int^{x}\sigma_{x}(x')dx'}.\label{eq:pml-attenuation}
\end{equation}
Notice the factor $k/\omega$, which is equal to $1/c_{x}$, the inverse
of the phase velocity $c_{x}$ in the $x$ direction. In a dispersionless
material (e.g. vacuum for light), $c_{x}$ is a constant independent
of velocity for a fixed angle, in which case the \emph{attenuation
rate} in the PML is \emph{independent} of frequency $\omega$: all
wavelengths decay at the same rate! In contrast, if we had left out
the $1/\omega$ then shorter wavelengths would decay faster than longer
wavelengths. On the other hand, the attenuation rate is \emph{not}
independent of the \emph{angle} of the light, a difficulty discussed
in Sec\@.~\ref{subsec:angle-dependence}.

\subsection{Truncating the computational region}

Once we have performed the PML transformation (\ref{eq:pml-transformation})
of our wave equations, the solutions are unchanged in our region of
interest (small $x$) and exponentially decaying in the outer regions
(large $x$). That means that we can \emph{truncate} the computational
region at some sufficiently large $x$, perhaps by putting a hard
wall (Dirichlet boundary condition). Because only the tiny exponential
tails ``see'' this hard wall and reflect off it, and even \emph{they}
attenuate \emph{on the way back} towards the region of interest, the
effect on the solutions in our region of interest will be exponentially
small. 

In fact, in principle we can make the PML region as thin as we want,
just by making $\sigma_{x}$ very large (which makes the exponential
decay rate rapid), thanks to the fact that the decay rate is independent
of $\omega$ (although the angle dependence can be a problem, as discussed
in Sec\@.~\ref{subsec:angle-dependence}). However, in practice,
we will see in Sec\@.~\ref{subsec:discretization} that using a
very large $\sigma_{x}$ can cause ``numerical reflections'' once
we discretize the problem onto a grid. Instead, we turn on $\sigma_{x}(x)$
quadratically or cubically from zero, over a region of length a half-wavelength
or so, and in practice the reflections will be tiny.

\subsection{PML boundaries in other directions}

So far, we've seen how to truncate our computational region with a
PML layer in the $+x$ direction. What about other directions? The
most important case to consider is the $-x$ direction. The key is,
in the $-x$ direction we do \emph{exactly the same thing}: apply
the PML transformation (\ref{eq:pml-transformation}) with $\sigma_{x}>0$
at a sufficiently large negative $x$, and then truncate the computational
cell. This works because, for $x<0$, the radiating waves are propagating
in the $-x$ direction with $k<0$ (negative phase velocity), and
this makes our PML solutions (\ref{eq:pml-attenuation}) decay in
the opposite direction (exponentially decaying as $x\rightarrow-\infty$)
for the \emph{same} positive $\sigma_{x}$.

Now that we have dealt with $\pm x$, the $\pm y$ and $\pm z$ directions
are easy: just do the same transformation, except to $\partial/\partial y$
and $\partial/\partial z$, respectively, using functions $\sigma_{y}(y)$
and $\sigma_{z}(z)$ that are non-zero in the $y$ and $z$ PML regions.
At the corners of the computational cell, we will have regions that
are PML along two or three directions simultaneously (i.e. two or
three $\sigma$'s are nonzero), but that doesn't change anything.

\subsection{Coordinate transformations and materials\label{subsec:coordinates-and-materials}}

We will see below that, in the context of the scalar wave equation,
the $1+i\sigma/\omega$ term from the PML coordinate transformation
appears as, effectively, an \emph{artificial anisotropic absorbing
material} in the wave equation (effectively changing $a$ and $b$
to complex numbers, and a tensor in the case of $a$). At least in
the case of Maxwell's equations (electromagnetism), this is an instance
of a more general theorem: Maxwell's equations under \emph{any coordinate
transformation} can be expressed as Maxwell's equations in Cartesian
coordinates with \emph{transformed materials}.\footnote{This theorem appears to have been first clearly stated and derived
by Ward and Pendry~\cite{Ward96}, and is summarized in a compact
general form by my course notes~\cite{Johnson07-coordinate-transforms}
and in our paper~\cite{KottkeFa08}.} That is, the coordinate transform is ``absorbed'' into a change
of the permittivity $\varepsilon$ and the permeability $\mu$ (generally
into anisotropic tensors). This is the reason why UPML, which constructs
reflectionless anisotropic absorbers, is equivalent to a complex coordinate
stretching: it is just absorbing the coordinate stretching into the
material tensors.

\section{PML examples in frequency and time domain}

As we have seen, in \emph{frequency domain}, when we are solving for
solutions with time-dependence $e^{-i\omega t}$, PML is almost trivial:
we just apply the PML transformation (\ref{eq:pml-transformation})
to every $\frac{\partial}{\partial x}$ derivative in our wave equation.
(And similarly for derivatives in other directions, to obtain PML
boundaries in different directions.)

In the time domain, however, things are a bit more complicated, because
we chose our transformation to be of the form $1+i\sigma/\omega$:
our complex ``stretch'' factor is frequency-\emph{dependent} in
order that the attenuation rate be frequency-\emph{independent}. But
how do we express a $1/\omega$ dependence in the time domain, where
we don't have $\omega$ (i.e. the time-domain wave function may superimpose
multiple frequencies at once)? One solution is to punt, of course,
and just use a stretch factor $1+i\sigma/\omega_{0}$ for some constant
frequency $\omega_{0}$ that is typical of our problem; as long as
our bandwidth is narrow enough, our attenuation rate (and thus the
truncation error) will be fairly constant. However, it turns out that
there is a way to implement the ideal $1/\omega$ dependence directly
in the time domain, via the \textbf{auxiliary differential equation}
(ADE) approach.

This approach is best illustrated by example, so we will consider
PML boundaries in the $x$ direction for the scalar wave equation
in one and two dimensions. (It turns out that an ADE is not required
in 1d, however.)

\subsection{An example: PML for 1d scalar waves}

Let's consider the 1d version of the scalar wave equation (\ref{eq:scalar-wave-udot}--\ref{eq:scalar-wave-vdot}):
\[
\frac{\partial u}{\partial t}=b\frac{\partial v}{\partial x}=-i\omega u
\]
\[
\frac{\partial v}{\partial t}=a\frac{\partial u}{\partial x}=-i\omega v,
\]
where we have substituted an $e^{-i\omega t}$ time-dependence. Now,
if we perform the PML transformation (\ref{eq:pml-transformation}),
and multiply both sides by $1+i\sigma_{x}/\omega$, we obtain: 
\[
b\frac{\partial v}{\partial x}=-i\omega u+\sigma_{x}u
\]
\[
a\frac{\partial u}{\partial x}=-i\omega v+\sigma_{x}v.
\]
The $1/\omega$ terms have cancelled, and so in this 1d case we can
trivially turn the equations back into their time-domain forms:
\[
\frac{\partial u}{\partial t}=b\frac{\partial v}{\partial x}-\sigma_{x}u
\]
\[
\frac{\partial v}{\partial t}=a\frac{\partial u}{\partial x}-\sigma_{x}v.
\]
Notice that, for $\sigma_{x}>0$, the decay terms have exactly the
right sign to make the solutions decay in \emph{time} if $u$ and
$v$ were constants in space. Similarly, they have the right sign
to make it decay in space whereever $\sigma_{x}>0$. But this is a
true PML: there are \emph{zero} reflections from any boundary where
we change $\sigma_{x}$, even if we change $\sigma_{x}$ discontinuously
(not including the discretization problems mentioned above).

By the way, the above equations reveal why we use the letter $\sigma$
for the PML absorption coefficient. If the above equations are interpreted
as the equations for electric ($u$) and magnetic ($v$) fields in
1d electromagnetism, then $\sigma$ plays the role of a \emph{conductivity},
and conductivity is traditionally denoted by $\sigma$. Unlike the
usual electrical conductivity, however, in PML we have both an electric
and a \emph{magnetic} conductivity, since we have terms corresponding
to currents of electric and magnetic charges. There is no reason we
need to be limited to \emph{physical} materials to construct our PML
for a computer \emph{simulation}!

\subsection{An example: PML for 2d scalar waves}

Unfortunately, the 1d case above is a little too trivial to give you
the full flavor of how PML works. So, let's go to a 2d scalar wave
equation (again for $e^{-i\omega t}$ time-dependence): 
\[
\frac{\partial u}{\partial t}=b\nabla\cdot v=b\frac{\partial v_{x}}{\partial x}+b\frac{\partial v_{y}}{\partial y}=-i\omega u
\]
\[
\frac{\partial v_{x}}{\partial t}=a\frac{\partial u}{\partial x}=-i\omega v_{x}
\]
\[
\frac{\partial v_{y}}{\partial t}=a\frac{\partial u}{\partial y}=-i\omega v_{y}.
\]
Again performing the PML transformation (\ref{eq:pml-transformation})
of $\frac{\partial}{\partial x}$ in the first two equations, and
multiplying both sides by $1+i\sigma_{x}/\omega$, we obtain: 
\[
b\frac{\partial v_{x}}{\partial x}+b\frac{\partial v_{y}}{\partial y}\left(1+i\frac{\sigma_{x}}{\omega}\right)=-i\omega u+\sigma_{x}u
\]
\[
a\frac{\partial u}{\partial x}=-i\omega v_{x}+\sigma_{x}v_{x}.
\]
The the second equation is easy to transform back to time domain,
just like for the 1d scalar-wave equation: $-i\omega$ becomes a time
derivative. The first equation, however, poses a problem: we have
an extra $\frac{ib\sigma_{x}}{\omega}\frac{\partial v_{y}}{\partial y}$
term with an explicit $1/\omega$ factor. What do we do with this?

In a Fourier transform, $-i\omega$ corresponds to differentiation,
so $i/\omega$ corresponds to \emph{integration}: our problematic
$1/\omega$ term is the \emph{integral} of another quantity. In particular,
let's introduce a new \emph{auxiliary} field variable $\psi$, satisfying
\[
-i\omega\psi=b\sigma_{x}\frac{\partial v_{y}}{\partial y},
\]
in which case
\[
b\frac{\partial v_{x}}{\partial x}+b\frac{\partial v_{y}}{\partial y}+\psi=-i\omega u+\sigma_{x}u.
\]
Now, we can Fourier transform everything back to the time-domain,
to get a set of \emph{four} time-domain equations with PML absorbing
boundaries in the $x$ direction that we can solve by our favorite
discretization scheme:
\[
\frac{\partial u}{\partial t}=b\nabla\cdot v-\sigma_{x}u+\psi
\]
 
\[
\frac{\partial v_{x}}{\partial t}=a\frac{\partial u}{\partial x}-\sigma_{x}v_{x}
\]
 
\[
\frac{\partial v_{y}}{\partial t}=a\frac{\partial u}{\partial y}
\]
\[
\frac{\partial\psi}{\partial t}=b\sigma_{x}\frac{\partial v_{y}}{\partial y},
\]
where the last equation for $\psi$ is our \emph{auxiliary differential
equation} (with initial condition $\psi=0$). Notice that we have
$\sigma_{x}$ absorption terms in the $u$ and $v_{x}$ equation,
but not for $v_{y}$: the PML corresponds to an \emph{anisotropic
absorber}, as if $a$ were replaced by the $2\times2$ complex tensor
\[
\left(\begin{array}{cc}
\frac{1}{a}+i\frac{\sigma_{x}}{\omega a}\\
 & \frac{1}{a}
\end{array}\right)^{-1}.
\]
This is an example of the general theorem alluded to in Sec\@.~\ref{subsec:coordinates-and-materials}
above. 

\section{PML in inhomogeneous media}

The derivation above didn't really depend at all on the assumption
that the medium was homogenous in $(y,z)$ for the $x$ PML layer.
We only assumed that the medium (and hence the wave equation) was
invariant in the $x$ direction for sufficiently large $x$. For example,
instead of empty space we could have a waveguide oriented in the $x$
direction (i.e. some $x$-invariant $yz$ cross-section). Regardless
of the $yz$ dependence, translational invariance implies that radiating
solutions can be decomposed into a sum of functions of the form of
eq\@.~(\ref{eq:planewave-x}), $\vec{W}(y,z)e^{i(kx-\omega t)}$.
These solutions $\vec{W}$ are no longer plane waves. Instead, they
are the \emph{normal modes} of the $x$-invariant structure, and $k$
is the \emph{propagation constant}. These normal modes are the subject
of waveguide theory in electromagnetism, a subject extensively treated
elsewhere \cite{Snyder70,Marcuse74}. The bottom line is: since the
solution/equation is still analytic in $x$, the PML is still reflectionless.\footnote{There is a subtlety here because, in unusual cases, uniform waveguides
can support ``backward-wave'' modes where the phase and group velocities
are opposite, i.e. $k<0$ for a right-traveling wave \cite{Clarricoats60,Waldron64,Omar87,Ibanescu04-anomalous}.
It has problems even worse than those reported for left-handed media
\cite{Cummer04,Dong04}, because the same frequency has both ``right-handed''
and ``left-handed'' modes; a deeper analysis of this interesting
case can be found in our paper~\cite{LohOs09}.}

\section{PML for evanescent waves}

In the discussion above, we considered waves of the form $e^{ikx}$
and showed that they became exponentially decaying if we replace $x$
by $x(1+i\sigma_{x}/\omega)$ for $\sigma>0$. However, this discussion
assumed that $k$ was \emph{real} (and positive). This is not necessarily
the case! In two or more dimensions, the wave equation can have \emph{evanescent}
solutions where $k$ is complex, most commonly where $k$ is purely
imaginary. For example consider a planewave $e^{i(\vec{k}\cdot\vec{x}-\omega t)}$
in a homogeneous two-dimensional medium with phase velocity $c$,
i.e. $\omega=c|\vec{k}|=c\sqrt{k_{x}^{2}+k_{y}^{2}}$. In this case,
\[
k=k_{x}=\sqrt{\frac{\omega^{2}}{c^{2}}-k_{y}^{2}}.
\]
For sufficiently large $k_{y}$ (i.e. high-frequency Fourier components
in the $y$ direction), $k$ is purely imaginary. As we go to large
$x$, the boundary condition at $x\rightarrow\infty$ implies that
we must have $\Im k>0$ so that $e^{ikx}$ is exponentially decaying.

What happens to such a decaying, imaginary-$k$ evanescent wave in
the PML medium? Let $k=i\kappa$. Then, in the PML: 
\begin{equation}
e^{-\kappa x}\rightarrow e^{-\kappa x-i\frac{\sigma_{x}}{\omega}x}.\label{eq:evanescent-pml-attenuation}
\end{equation}
That is, the PML added an oscillation to the evanescent wave, but
\emph{did not increase its decay rate}. The PML is still reflectionless,
but it didn't \emph{help}.

Of course, you might object that an evanescent wave is decaying \emph{anyway},
so we hardly need a PML---we just need to make the computational
region large enough and it will go away on its own. This is true,
but it would be nice to accelerate the process: in some cases $\kappa=\Im k$
may be relatively small and we would need a large grid for it to decay
sufficiently. This is no problem, however, because \emph{nothing}
in our analysis required $\sigma_{x}$ to be \emph{real}. We can just
as easily make $\sigma_{x}$ \emph{complex}, where $\Im\sigma_{x}<0$
corresponds to a \emph{real} coordinate stretching. That is, the imaginary
part of $\sigma_{x}$ will accelerate the decay of evanescent waves
in eq\@.~(\ref{eq:evanescent-pml-attenuation}) above, without creating
any reflections.

Adding an imaginary part to $\sigma_{x}$ does come at a price, however.
What it does to the \emph{propagating} (real $k$) waves is to make
them oscillate \emph{faster}, which exacerbates the numerical reflections
described in Sec\@.~\ref{subsec:discretization}. In short, everything
in moderation.

\section{Limitations of PML\label{sec:Limitations-of-PML}}

PML, while it has revolutionized absorbing boundaries for wave equations,
especially (but not limited to) electromagnetism, is not a panacea.
Some of the limitations and failure cases of PML are discussed in
this section, along with workarounds.

\subsection{Discretization and numerical reflections\label{subsec:discretization}}

First, and most famously, PML is only reflectionless if you are solving
the \emph{exact} wave equations. As soon as you discretize the problem
(whether for finite difference or finite elements), you are only solving
an approximate wave equation and the analytical perfection of PML
is no longer valid.

What is left, once you discretize? PML is still an absorbing material:
waves that propagate within it are still attenuated, even discrete
waves. The boundary between the PML and the regular medium is no longer
reflectionless, but the reflections are small because the discretization
is (presumably) a good approximation for the exact wave equation.
How can we make the reflections smaller, as small as we want?

The key fact is that, even without a PML, reflections can be made
arbitrarily small as long as the medium is \emph{slowly varying}.
That is, in the limit as you ``turn on'' absorption more and more
slowly, reflections go to zero due to an \emph{adiabatic theorem}
\cite{Johnson02:adiabatic}. With a non-PML absorber, you might need
to go \emph{very} slowly (i.e. a very thick absorbing layer) to get
acceptable reflections \cite{OskooiZh08}. With PML, however, the
constant factor is very good to start with, so experience shows that
a simple quadratic or cubic turn-on of the PML absorption usually
produces negligible reflections for a PML layer of only half a wavelength
or thinner \cite{Taflove00,OskooiZh08}. (Increasing the resolution
also increases the effectiveness of the PML, because it approaches
the exact wave equation.) 

\subsection{Angle-dependent absorption\label{subsec:angle-dependence}}

Another problem is that the PML absorption depends on angle. In particular,
consider eq\@.~(\ref{eq:pml-attenuation}) for the exponential attenuation
of waves in the PML, and notice that the attenuation rate is proportional
to the ratio $k/\omega$. But $k$, here, is really just $k_{x}$,
the \emph{component} of the wavevector $\vec{k}$ in the $x$ direction
(for a planewave in a homogeneous medium). Thus, the attenuation rate
is proportional to $|\vec{k}|\cos\theta$, where $\theta$ is the
angle the radiating wave makes with the $x$ axis. As the radiation
approaches glancing incidence, the attenuation rate goes to zero!
This means that, for any fixed PML thickness, waves sufficiently close
to glancing incidence will have substantial ``round-trip'' reflections
through the PML. 

In practice, this is not as much of a problem as it may sound like
at first. In most cases, all of the radiation originates in a localized
region of interest near the origin, as in Fig\@.~\ref{fig:absorbing-layer}.
In this situation, all of the radiation striking the PML will be at
an angle $\theta<55^{\circ}\approx\cos^{-1}(1/\sqrt{3})$ in the limit
as the boundaries move farther and farther away (assuming a cubic
computational region). So, if the boundaries are far enough away,
we can guarantee a maximum angle and hence make the PML thick enough
to sufficiently absorb all waves within this cone of angles.

\subsection{Inhomogeneous media where PML fails\label{subsec:inhomogeneous-failure}}

Finally, PML fails completely in the case where the medium is not
$x$-invariant (for an $x$ boundary) \cite{OskooiZh08}. You might
ask: why should we care about such cases, as if the medium is varying
in the $x$ direction then we will surely get reflections (from the
variation) anyway, PML or no PML? Not necessarily.

There are several important cases of $x$-varying media that, in the
infinite system, have reflectionless propagating waves. Perhaps the
simplest is a waveguide that hits the boundary of the computational
cell at an angle (not normal to the boundary)---one can \emph{usually}
arrange for all waveguides to leave the computational region at right
angles, but not \emph{always} (e.g. what if you want the transmission
through a $30^{\circ}$ bend?). Another, more complicated and perhaps
more challenging case is that of a \emph{photonic crystal}: for a
\emph{periodic} medium, there are wave solutions (\emph{Bloch waves})
that propagate without scattering, and can have very interesting properties
that are unattainable in a physical uniform medium \cite{Joannopoulos95}.

For any such case, PML seems to be irrevocably spoiled. The central
idea behind PML was that the wave equations, and solutions, were analytic
functions in the direction perpendicular to the boundary, and so they
could be analytically continued into the complex coordinate plane.
If the medium is varying in the $x$ direction, it is most likely
varying discontinously, and hence the whole idea of analytic continuation
goes out the window.

What can we do in such a case? Conventional ABCs don't work either
(they are typically designed for homogeneous media). The only fallback
is the adiabatic theorem alluded to above: even a non-PML absorber,
if turned on gradually enough and smoothly enough, will approach a
reflectionless limit. The difficulty becomes how gradual is gradual
enough, and in finding a way to make the non-PML absorber a tractable
thickness \cite{OskooiZh08}.

There is also another interesting case where PML fails. The basic
analytic-continuation idea is valid in any $x$-invariant medium,
regardless of inhomogeneities in the $yz$ plane. However, certain
inhomogeneous dielectric patterns in the $yz$ plane give rise to
unusual solutions: \textquotedblleft left-handed\textquotedblright{}
solutions where the phase velocity is opposite to the group velocity
in the $x$ direction, while at the same $\omega$ the medium \emph{also
}has \textquotedblleft right-handed\textquotedblright{} solutions
where the phase velocity and group velocity have the same sign. Most
famously, this occurs for certain ``backward-wave'' coaxial waveguides~\cite{Clarricoats60,Waldron64,Omar87,Ibanescu04-anomalous}.
In this case, whatever sign one picks for the PML conductivity $\sigma$,
either the left- or right-handed modes will be exponentially growing
and the PML fails in a spectacular instability~\cite{LohOs09}. There
is a subtle relationship of this failure to the orientation of the
fields for a left-handed mode and the anisotropy of the PML~\cite{LohOs09}.
In this case, one must once again abandon PML absorbers and use a
different technique, such as a scalar conductivity that is turned
on sufficiently gradually to adiabatically absorb outgoing waves.

\bibliographystyle{ieeetr}
\bibliography{teaching}

\end{document}